\documentclass[]{aa}
\usepackage{graphicx}
\usepackage{natbib}
\usepackage{amssymb}
\usepackage{longtable}
\usepackage{amsmath}
 
\bibpunct{(}{)}{;}{a}{}{,}     % A&A citation style
\newcommand{\micron}{$\mu$m}
\newcommand{\ra}[4]{$#1^{\mathrm{h}}\,#2^{\mathrm{m}}\,#3^{\mathrm{s}}.#4$}
\newcommand{\dec}[4]{$#1\degr\,#2'\,#3''.#4$}
\title{Lack of PAH emission toward low-mass embedded young stellar objects\thanks{Based on
observations obtained at the European Southern Observatory, Paranal,
Chile, within the observing program 164.I-0605 (VLT-ISAAC).}}
\titlerunning{Lack of PAH emission toward low-mass embedded YSOs}

\author{
V.\ C. Geers \inst{1,2}
\and E.\ F. van Dishoeck\inst{1}
\and K.\ M. Pontoppidan\inst{3}
\and F.\ Lahuis\inst{4}
\and A.\ Crapsi\inst{1,5}
\and C.\ P.\ Dullemond\inst{6}
\and G.\ A.\ Blake\inst{3}
}

\offprints{V.\,C. Geers,\\ \email{vcgeers@astro.utoronto.ca}}

\institute{Leiden Observatory, Leiden University, P.O.\ Box 9513, 2300 RA Leiden, The Netherlands
\and University of Toronto, 50 St.\ George St., Toronto, ON M5R 2W9, Canada   
\and Division of Geological and Planetary Sciences, Mail Code 150-21, California Institute of Technology, Pasadena, CA 91125, USA
\and SRON Netherlands Institute for Space Research, P.O.\ Box 800, 9700 AV Groningen, The Netherlands
\and Observatorio Astron\'{o}mico Nacional (IGN), Alfonso XII, 3, E-28014 Madrid, Spain
\and Max-Planck-Institut f\"{u}r Astronomie, Koenigstuhl 17, 69117 Heidelberg, Germany
}
\authorrunning{Geers et al.}

\date{Received 19 September 2008; Accepted 2 December 2008}
\abstract
{}
{Polycyclic Aromatic Hydrocarbons (PAHs) have been detected toward molecular clouds and some young stars with disks, but have not yet been associated with embedded young stars. We present a sensitive mid-infrared spectroscopic survey of PAH features toward a sample of low-mass embedded young stellar objects (YSOs). The aim is to put constraints on the PAH abundance in the embedded phase of star formation using radiative transfer modeling.}
{VLT-ISAAC L-band spectra for 39 sources and Spitzer IRS spectra for 53 sources are presented. Line intensities are compared to recent surveys of Herbig Ae/Be and T Tauri stars. The radiative transfer codes RADMC and RADICAL are used to model the PAH emission from embedded YSOs consisting of a pre-main-sequence star with a circumstellar disk embedded in an envelope. The dependence of the PAH feature on PAH abundance, stellar radiation field, inclination and the extinction by the surrounding envelope is studied.}
{The 3.3 \micron\ PAH feature is undetected for the majority of the sample (97\%), with typical upper limits of 5 $\times 10^{-16}$ W m$^{-2}$. One source originally classified as class I, IRS~48, shows a strong 3.3 \micron\ feature from a disk. Compact 11.2 \micron\ PAH emission is seen directly towards 1 out of the 53 Spitzer Short-High spectra, for a source that is borderline embedded. For all 12 sources with both VLT and Spitzer spectra, no PAH features are detected in either. In total, PAH features are detected toward at most 1 out of 63 (candidate) embedded protostars ($\lesssim$ 2\%), even lower than observed for class II T Tauri stars with disks (11--14\%). Models predict the 7.7 \micron\ feature as the best tracer of PAH emission, while the 3.3 \micron\ feature is relatively weak. Assuming typical class I stellar and envelope parameters, the absence of PAHs emission is most likely explained by the absence of emitting carriers through a PAH abundance at least an order of magnitude lower than in molecular clouds but similar to that found in disks. Thus, most PAHs likely enter the protoplanetary disks frozen out in icy layers on dust grains and/or in coagulated form.
}
{}


\begin{document}
\maketitle

%############
\section{Introduction}
Polycyclic Aromatic Hydrocarbons (PAHs) have been observed toward a wide range of astrophysical environments \citep{all89,pee02}, including the interstellar medium (ISM) and star-forming regions,
 first hinted at by the discovery of widespread broad emission features in the mid-infrared.
Since PAHs are a good tracer of UV radiation, they are an indirect probe of star formation in the high opacity environments of molecular clouds as well as circumstellar disks. Recent spectroscopic studies have detected PAHs toward a significant fraction, 11--14\% and 54\% of low-mass and intermediate mass pre-main-sequence (PMS) stars respectively (\citealt{ack04}; \citealt[][chapter 2]{gee07c}), and recent ground-based high spatial resolution observations show evidence for the PAH emission to originate from the circumstellar disks \citep[]{res03,hab06,gee07,gee07b}.
PAHs are also prominently seen toward photon dominated regions (PDRs) in dense clouds exposed to massive young stars \citep[e.g.,\ ][]{ver96}. However, to this date, no significant PAH emission has been reported to be directly associated with the earlier class 0--I phase, when a (proto)star with circumstellar disk is still embedded in an envelope of gas and dust. Deeply embedded high-mass young stellar objects (YSO's) also lack prominent PAH emission \citep{dis00}.

PAHs are believed to form in the outflows of carbon-rich Asymptotic Giant Branch (AGB) stars, which deposit them in the interstellar medium \citep[e.g.,\ ][]{hel96}. Their ubiquitous presence in the ISM shows that at least larger PAH molecules (of 50--100 carbon atoms or more) will survive this phase, until being incorporated in molecular clouds. The amount of carbon locked up in PAHs is [C/H]$_{\mathrm{PAH}} \simeq 5 \times 10^{-5}$ \citep{hab04}, corresponding PAH abundances of $\sim$~$5 \times 10^{-7}$ relative to H, for PAH molecules of 100 carbon atoms, making them among the most abundant molecules after H$_2$ and CO. 

PAHs can play an important role in the star-forming environments. As large molecules, they provide efficient heating of the gas in both the ISM and circumstellar disks. Small dust particles like PAHs have also been suggested as the main formation site of molecules such as H$_2$ and water in the later evolutionary phases of circumstellar disks when classical grains have grown to large sizes \citep{jon06}. 

The shape and relative strength of the PAH features in the 6--9 \micron\ region has been shown to vary between various astrophysical environments from AGBs, the ISM to the class II disk sources. This has provided evidence that the emission characteristics of PAHs are sensitive to local physical conditions and  that interstellar PAHs undergo processing in space environments \citep{pee02}. Further support for this comes from recent values for the PAH abundance in disk surface layers, which is a factor of  10-100 lower than in the ISM for sources with sufficient UV radiation to excite them \citep{gee06}. PAHs frozen out on icy grains can chemically react with other molecules to form a large variety of complex species, including pre-biotic molecules \citep{ber99,ehr06}.
A study of PAHs toward embedded low-mass young stars is important to find out what happens to PAHs and how large a role they may play in these embedded objects.

In this paper we present the first mid-infrared spectroscopic survey for PAH emission from embedded low-mass young stellar objects using the Infrared Spectrometer And Array Camera (ISAAC) mounted on the ESO Very Large Telescope (VLT) and the Infrared Spectrograph (IRS) on board the NASA Spitzer Telescope, in a wide range of nearby star-forming regions. We subsequently use a radiative transfer code to model PAH emission from young stars with disks embedded in an envelope, and discuss the observed detection rate in the context of PAH abundance, the strength of UV emission from the protostar  and the extinction from the envelope.

%############
\section{Observations and data reduction}

A large sample of low-mass embedded objects was selected from two previous mid-infrared studies. A set of 39 sources was selected from a VLT-ISAAC 3--5 \micron\ band survey \citep{dis03}, of which 32 were previously presented in \citet{pon03} and three additional sources in \citet{thi06}. For this survey L-band spectroscopy was obtained with ISAAC, the Infrared Spectrometer And Array Camera, installed at the VLT Antu (UT1) at ESO's Paranal Observatory in Chile, in the low resolution ($R = \lambda / \Delta \lambda$ = 600--1200) spectroscopic mode in the spectral domain 2.8--4.2 $\mu$m using a $0.6\arcsec\times120\arcsec$ slit. The telescope was operated using a standard chop-throw scheme with typical chop-throws of 10--20\arcsec. A full description of observation and reduction techniques is given in \citet{pon03}. These sources were required to have a rising Spectral Energy Distribution (SED) in the mid-infrared as well as the presence of an H$_{2}$O ice feature at 3 \micron. The selected sources are listed in Table\,\ref{embeddedpah:tab:obssum}. 

In addition, a sample of 53 Spitzer Space Telescope Infrared Spectrograph (Spitzer IRS) Short-High (10--20 \micron, $R\sim$600) spectra and 33 Short-Low (5--14.5 \micron, $R\sim$100) of class I sources were obtained in the context of the ``Cores to Disks'' (c2d) Legacy program \citep{eva03}.
These sources were selected for showing the silicate 10 \micron\ feature in absorption, and the list includes a small number of known or candidate edge-on disks, which are labelled in Table\,\ref{embeddedpah:tab:obssum}. Most of them have spectral slopes $\alpha$ between 2 and 24 \micron\ $>$ 0. The vast majority of class 0 and I sources in the c2d sample were found to have luminosities of $L_{\mathrm{bol}} < 3$ L$_{\odot}$ \citep{eva08} indicative of low mass young stars and consistent with the observational fact that the majority of the young stars in the regions studied by c2d are K- and M-type stars \citep{luh07,oli08}. The reduction of the Spitzer spectroscopy was performed using the optimal PSF extraction technique developed within the c2d program, to allow separation of the compact source and extended emission components \citep{lah07}. The source size is determined from the width of the PSF function fitted to the source, compared to the width of the PSF function fit for standard calibrator stars. Comparison of the optimal PSF extraction and the full aperture extraction provides a direct estimate of any potential extended emission. An application of the same technique to remove background PAH features has been given for the source VSSG~1 in \citet[see their Fig.\ 4]{gee06}. 
The details of the observations and reduction procedures are described in \citet{lah06} and Lahuis (2007, thesis Chap.\ 3).

In total, the sample contains 80 sources, for which 12 sources were observed with both ISAAC and Spitzer. Based on new 2MASS-Spitzer-MIPS 2--24 \micron\ spectral slopes and/or sub-mm line and continuum data, as well as CO 4.7 \micron\ spectroscopy, the classification has been refined (\citealt{kem08,eno08,kem08t,jor08}, Pontoppidan priv.\ comm.). These updates have confirmed 53 embedded sources and 17 sources to be disk sources, while for 10 sources the classification remains uncertain (Table\,\ref{embeddedpah:tab:obssum}).

%############
\section{Results and discussion}
\subsection{VLT-ISAAC spectra}
\begin{figure}
  \centering
  \includegraphics[width=\columnwidth]{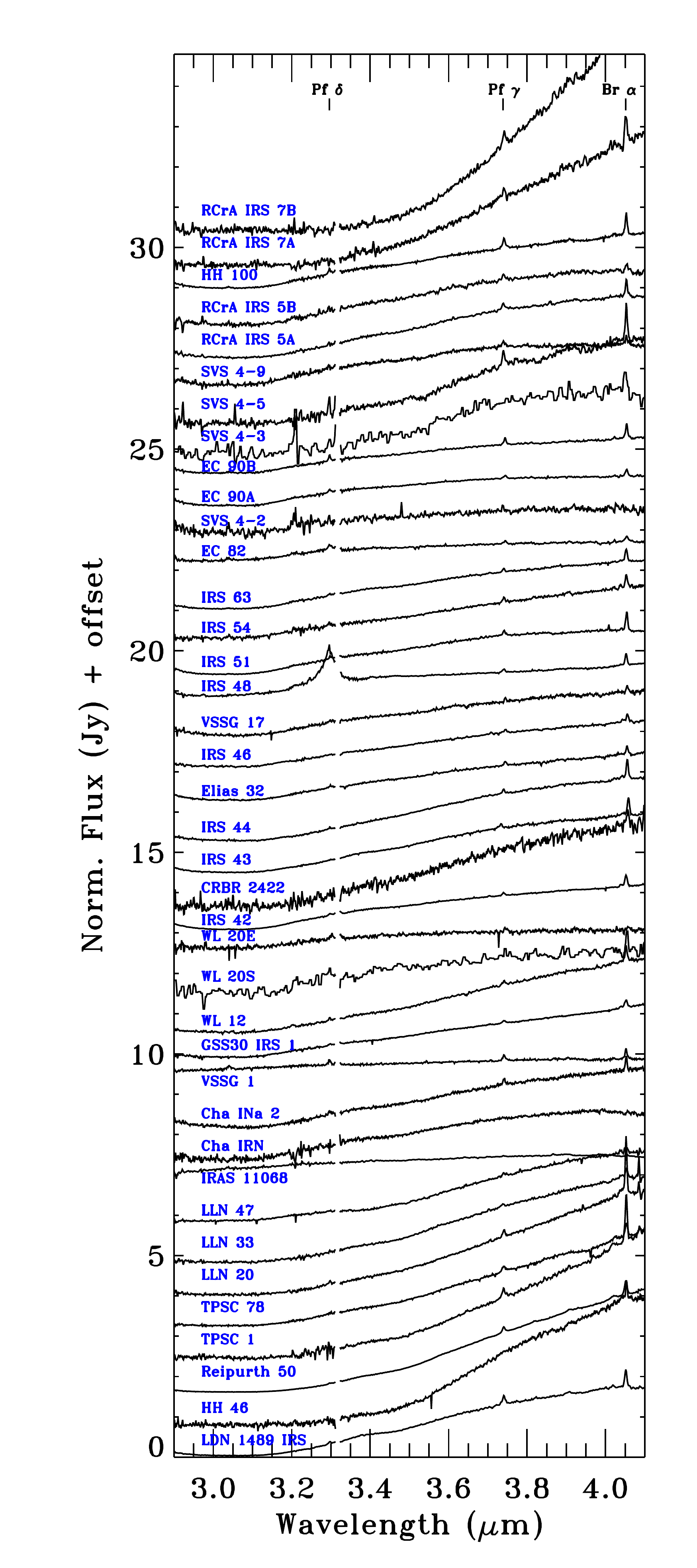}
  \caption{ISAAC L-band spectra of low-mass embedded sources. The narrow HI lines at 3.3, 3.74 and 4.05 \micron\ are seen in most spectra.}
  \label{embpah:fig:isaac1}
\end{figure}

The VLT-ISAAC L-band spectra are presented in Fig.~\ref{embpah:fig:isaac1}. 
Only 1 source in this sample shows a clear 3.3 \micron\ PAH emission feature, IRS~48. This source is known to have a spatially resolved disk surrounded by a very tenuous envelope at most \citep{gee07}. No clear PAH feature is seen toward any of the other sources. 
To derive upper limits, a gaussian with a peak flux equal to the 3$\sigma$ noise and FWHM = 0.07 \micron\ is generated and integrated to retrieve the line flux of the largest PAH 3.3 \micron\ feature that would still fall within the noise. The results are listed in Table~\ref{embeddedpah:tab:obssum}.

The hydrogen Pf $\delta$ (3.296 \micron), Pf $\gamma$ (3.739 \micron) and Br $\alpha$ (4.051 \micron) are detected toward most sources, with the exception of IRAS~11068-7717 and Cha~IRN. These lines are believed to be produced in the accretion column onto the central star \citep{cal98}, and/or in the wind launched close to the star \citep{kra08}. The broad absorption features by water ice (2.8 \micron) and the 3.47 \micron\ ice feature can be clearly seen toward the majority of the sources.

\subsection{Spitzer spectra}
\begin{figure}
  \centering
  \includegraphics[width=\columnwidth]{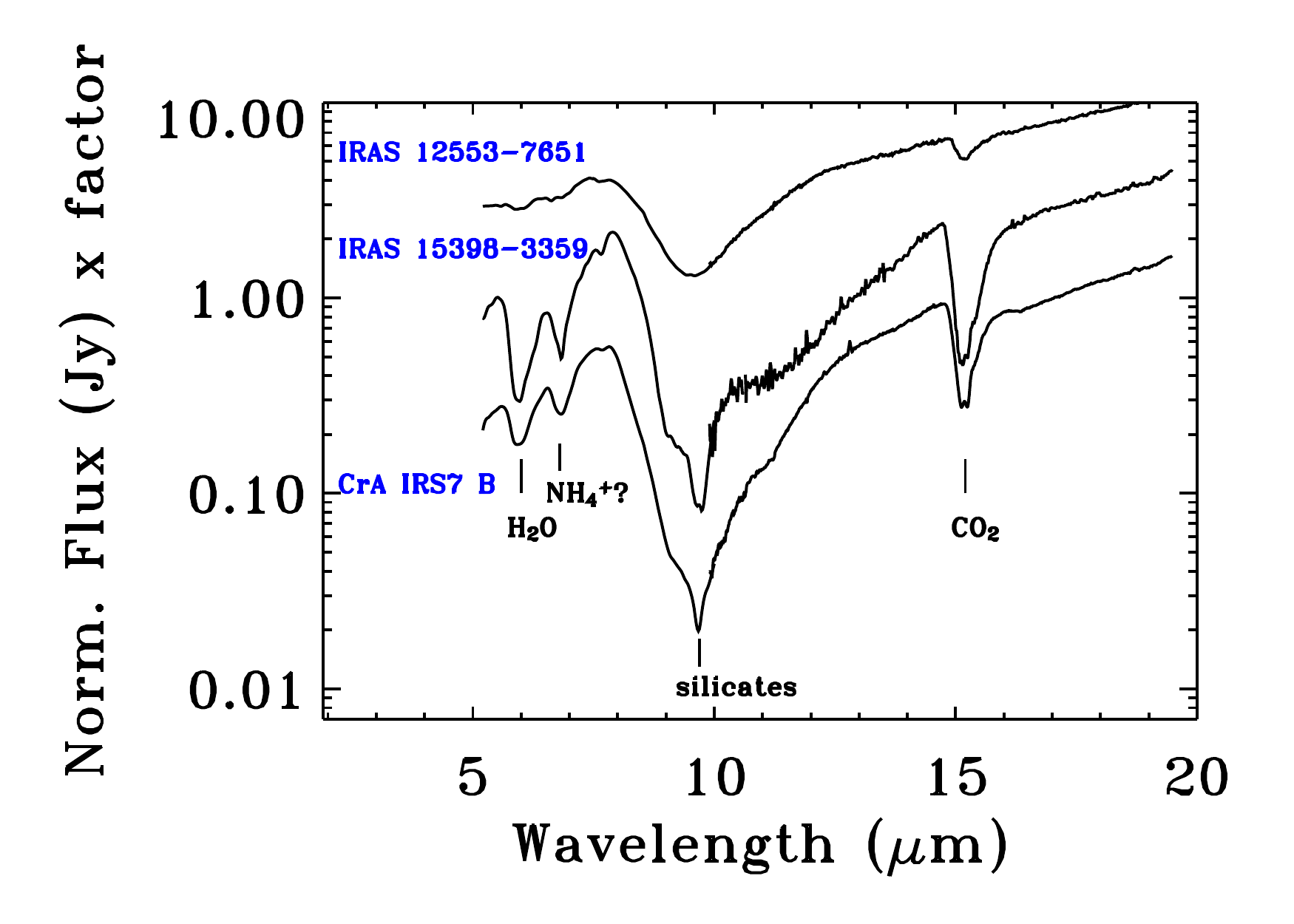}
  \caption{Spitzer IRS spectra of a sample of low-mass embedded sources without PAH detections. For clarity, the spectra are normalized to their flux at 17 \micron\ and scaled by factors of 1, 3 and 8 for CrA~IRS7~B, IRAS15398-3359 and IRAS~12553-7651 respectively.}
  \label{embpah:fig:spitzer1}
\end{figure}

Out of the sample of 53 sources, 50 do not show any PAH features. A selection of Spitzer spectra with PAH non-detections is presented in Fig.~\ref{embpah:fig:spitzer1}. Silicate, water and CO$_2$ absorption are the predominant features in these spectra. These have been analyzed by \citet{boo08} and \citet{pon08}.

\begin{figure}
  \centering
  \includegraphics[width=\columnwidth]{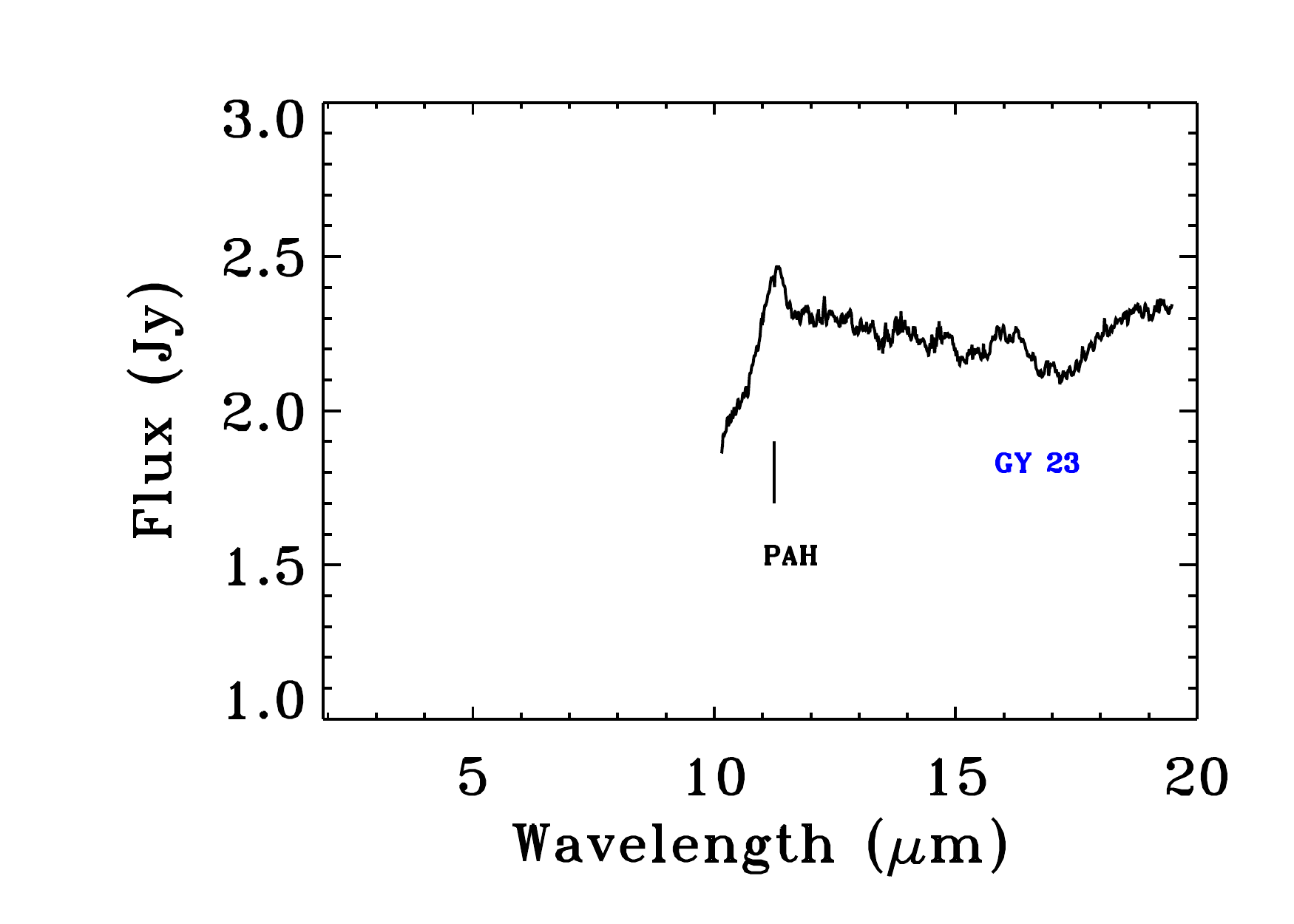}
  \caption{Spitzer IRS spectra of GY~23, a low-mass borderline embedded source with a PAH detection.}
  \label{embpah:fig:spitzer2}
\end{figure}

For one Spitzer source, Oph~GY~23, the optimal extraction method produces a SH spectrum with the observed 11.2 \micron\ PAH features, shown in Fig.\ \ref{embpah:fig:spitzer2}. 
A small size is derived for this source, suggesting that the PAH emission is directly associated with the star. The classification of this source remains uncertain. Based on Spitzer IRS, IRAC and MIPS fluxes, Oph~GY~23, previously classified as class I, shows a declining SED suggestive of a class II source. However, GY~23 is surrounded by reflection nebulosity and deep hot CO gas absorption features have been observed towards this source (Pontoppidan, priv. comm.), hinting at the presence of remnant envelope material. 

For two sources, VSSG~1 and IRS~14 in Ophiuchus, PAH features are detected from the extended background emission, which is assumed not to be directly associated with a central object.

Overall, no PAHs are detected toward any of the 53 confirmed embedded sources. PAHs are detected towards one of the 17 confirmed disk sources, IRS~48, and one out of the 10 sources with an uncertain classification, GY~23. 
%

%############
\section{Radiative transfer model}
Here we address the question of where the PAH emission can arise in a disk + envelope system through radiative transfer modeling and what the non-detections imply quantitatively about the PAH abundance.
\subsection{Physical structure}
We use the two-dimensional axisymmetric radiative transfer code RADMC
\citep{dul04} to calculate the temperature structure and scattering source function for an embedded YSO using a Monte Carlo technique. This code requires stellar parameters, a stellar radiation field, a density structure and a set of dust opacities. Scattering is assumed to be isotropic. A module to treat the emission from quantum-heated PAH molecules and Very Small Grains, previously described in \citet{gee06}, has been included. To generate images and spectra, ray-tracing is performed using RADICAL \citep{dul00}. All spectra and SEDs presented here are scaled to a distance to the observer of 150 pc. 

The density structure adopted here follows \citet{cra08} and is comprised of three components: the disk, the envelope and the outflow cone. The adopted density structure for the disk has a power-law dependence along the radial coordinate and a gaussian dependence in height, and can be expressed as
\begin{equation*}
\centering
\rho_{\mathrm{disk}}(r,\theta) = \frac{\Sigma_{0} (r/R_{0})^{-1}}{\sqrt{2\pi}H(r)} \exp \left \{-\frac{1}{2} \left [\frac{r \cos \theta}{H(r)}\right ]^{2}\right \},
\end{equation*}
where $\theta$ is the angle from the axis of symmetry. The variation of scale-height, i.e., the flaring of the disk, is described in the function $H(r)=r \cdot H_0/R_0 \cdot (r/R_0)^{2/7}$, corresponding to the self-irradiated passive disk of \citet{cha97}. For the disk in our template model, we fix the outer radius $R_0$ = 300 AU, inner radius $R_{\mathrm{in}}=0.1$ AU, $H_0$ = 90 AU, and the disk mass to $M_{\mathrm{disk}} = 5\times10^{-3}$ M$_{\odot}$, see also Table \ref{embpah:tab:modelparams}. 

The envelope density follows the theoretical structure for a rotating and collapsing spheroid as derived by \citet{ulr76}, defined by\\
% referee format
%\begin{equation*}
%\centering
%\rho_{\mathrm{env}}(r,\theta) = \rho_0 \left (\frac{R_{\mathrm{rot}}}{r}\right )^{1.5}\left (1+ \frac{\cos \theta}{\cos \theta_0}\right )^{-0.5}\left (\frac{\cos \theta}{2 \cos \theta_0} + \frac{R_{\mathrm{rot}}}{r} \cos^2 \theta_0 \right )^{-1},
%\end{equation*}
% printer format
\begin{equation*}
\centering
\scriptsize
\begin{split}
\rho_{\mathrm{env}}(r,\theta) = \rho_0 \left (\frac{R_{\mathrm{rot}}}{r}\right )^{1.5}\left (1+ \frac{\cos \theta}{\cos \theta_0}\right )^{-0.5}\left (\frac{\cos \theta}{2 \cos \theta_0} + \frac{R_{\mathrm{rot}}}{r} \cos^2 \theta_0 \right )^{-1},
\end{split} 
\end{equation*}
where $\theta_0$ is the solution of the parabolic motion of an infalling particle which is given by $r/R_{\mathrm{rot}} \cdot (\cos \theta_0 - \cos \theta)/(\cos \theta_0 \sin^2 \theta_0) = 1$, $R_{\mathrm{rot}}$ is the centrifugal radius of the envelope, and $\rho_0$ is the density of the equatorial plane at the centrifugal radius. The outer radius of the envelope is fixed to 10\,000 AU and the centrifugal radius is set to 300 AU. The value of $\rho_0$ is varied to cover a range of envelope mass $M_{\mathrm{env}}$ between 0.1 and 1.5 M$_{\odot}$, the template model has $M_{\mathrm{env}}$ = 1.0 M$_{\odot}$. This is somewhat more massive than the average Class I source, which has typically M$_{\mathrm{env}} < 0.5$ M$_{\odot}$. The inner radius of the envelope is determined by the outflow cavity, which crosses the equatorial plane at $\sim$20 AU. The streamline outflow is included by setting the density of the regions where $\cos \theta_0$ is larger than $\cos 15\degr$ to the same value of the density of the envelope at the outer radius. This results in a funnel-shaped cavity which is conical only at large scales, where it presents a semi-aperture of 15$\degr$. The envelope introduces an optical extinction of A$_{\mathrm{V}} = 33$ at $i = 45 \degr$, for the template model with $M_{\mathrm{env}}$ = 1.0 M$_{\odot}$.

For the radiation field, a blackbody spectrum is taken with $T_{\mathrm{eff}}$ = 4000 K and a luminosity of 1 L$_{\odot}$ for the template model. In addition, given the strong dependence of PAHs on UV excitation and observations indicating the presence of UV excess around actively accreting young low-mass PMS stars \citep{bas89,ken90}, the radiation field for $\lambda < 0.8$ \micron\ is substituted by a modified Draine field \citep{dis82}, scaled up to ensure continuous overlap at $\lambda > 0.8$ \micron. The resulting spectrum is normalized to a luminosity of 1 L$_{\odot}$. For comparison, a separate model is run without the modified radiation field. Any radiation field produced by a higher mass (A-type) star would be in between these two extremes. The excess Draine UV field corresponds to an intensity relative to the mean interstellar radiation field, $G_0$, of $8 \times 10^4$ at a radius of 100 AU at the surface of the disk. 
\begin{table}
\centering
\caption{Parameters of the template model.}
\label{embpah:tab:modelparams}
\begin{tabular}{ll}
\hline
\hline
Parameter & value\\
\hline
Radiation field & 4000 K blackbody\\
 &  + excess UV field\\
$L_{*}$ & 1 L$_{\odot}$\\
$M_{*}$ & 1.0 M$_{\odot}$\\
$M_{\mathrm{disk}}$ & $5 \times 10^{-2}$ M$_{\odot}$ \\
$R_{\mathrm{disk,in}}$ & 0.1 AU\\
$R_{\mathrm{disk,out}}$ & 300 AU\\
$M_{\mathrm{env}}$ & 1.0 M$_{\odot}$ \\
$R_{\mathrm{env,out}}$ & 10000 AU\\
$R_{\mathrm{rot}}$ & 300 AU\\
$\theta_0$ & 15\degr\\
PAH abun. & 5$\times 10^{-7}$ w.r.t.\ H\\
\hline
\end{tabular}
\end{table}

\subsection{Treatment of dust and PAHs}
For the optical properties of the dust grain population, we adopt the set of dust mixtures and opacities used in \citet{cra08}: a mixture of 71\% silicates with a size distribution similar to that of \citep{wei01} with $\alpha_{\mathrm{WD}}$-parameter equal to -2 and turnover point for radii larger than 0.3 \micron\ covered by a mantle of ices with a water ice abundance of $3.0 \times 10^{-4}$ relative to H$_2$, together with 29\% of carbonaceous grains with a size distribution similar to that of \citet{wei01}, characterized by a turnover point at radius of 4.5 \micron, an $\alpha_{\mathrm{WD}}$-parameter of -2 and a water ice abundance of $1.5 \times 10^{-5}$ relative to H$_2$. A second dust population is added without the ice layer, for regions where the temperature is higher than 90 K, when ices will have evaporated. These are included by running the Monte Carlo code once with ices in the whole grid to calculate the temperature structure, and then using this information to replace the dust opacities where the temperature is higher than 90 K. The ice features include H$_2$O, CO$_2$ and CO, but not more complex species which could be responsible for (part of) the observed 6.0 and 6.8 \micron\ ice features \citep{boo08}.

PAHs are added as a dust species to the disk and the envelope. The PAH emission is calculated for an equal mix of neutral and singly ionized C$_{100}$H$_{24}$ molecules, adopting the
\citet{dra01} PAH emission model, using the ``thermal continuous''
approximation. Multi-photon events are included for the PAH
excitation, following the method outlined by
\citet{sie02}. 
We include the opacities from \citet{mat05a} for near-infrared wavelengths. Model calculations by \citet{vis07} show that PAHs with $N_{\mathrm{c}} = 100$ have longer lifetimes than the lifetime of the disk, and we can therefore safely discard the possibility of PAH destruction and keep the PAHs at constant abundance throughout the entire disk.

In our template model, the total PAH abundance is the same as used in \citet{gee06}, as a mass fraction of the dust of 0.061, which corresponds to a carbon abundance of 5$\times 10^{-5}$ with respect to hydrogen and a $N_{\mathrm{c}} = 100$ PAH abundance of 5$\times 10^{-7}$ with respect to hydrogen. This PAH abundance is divided into 50\% ionized and 50\% neutral PAHs, the same as in our previous disk-only models \citep{gee06,gee07b}.

\begin{figure}
  \centering
  \includegraphics[width=\columnwidth]{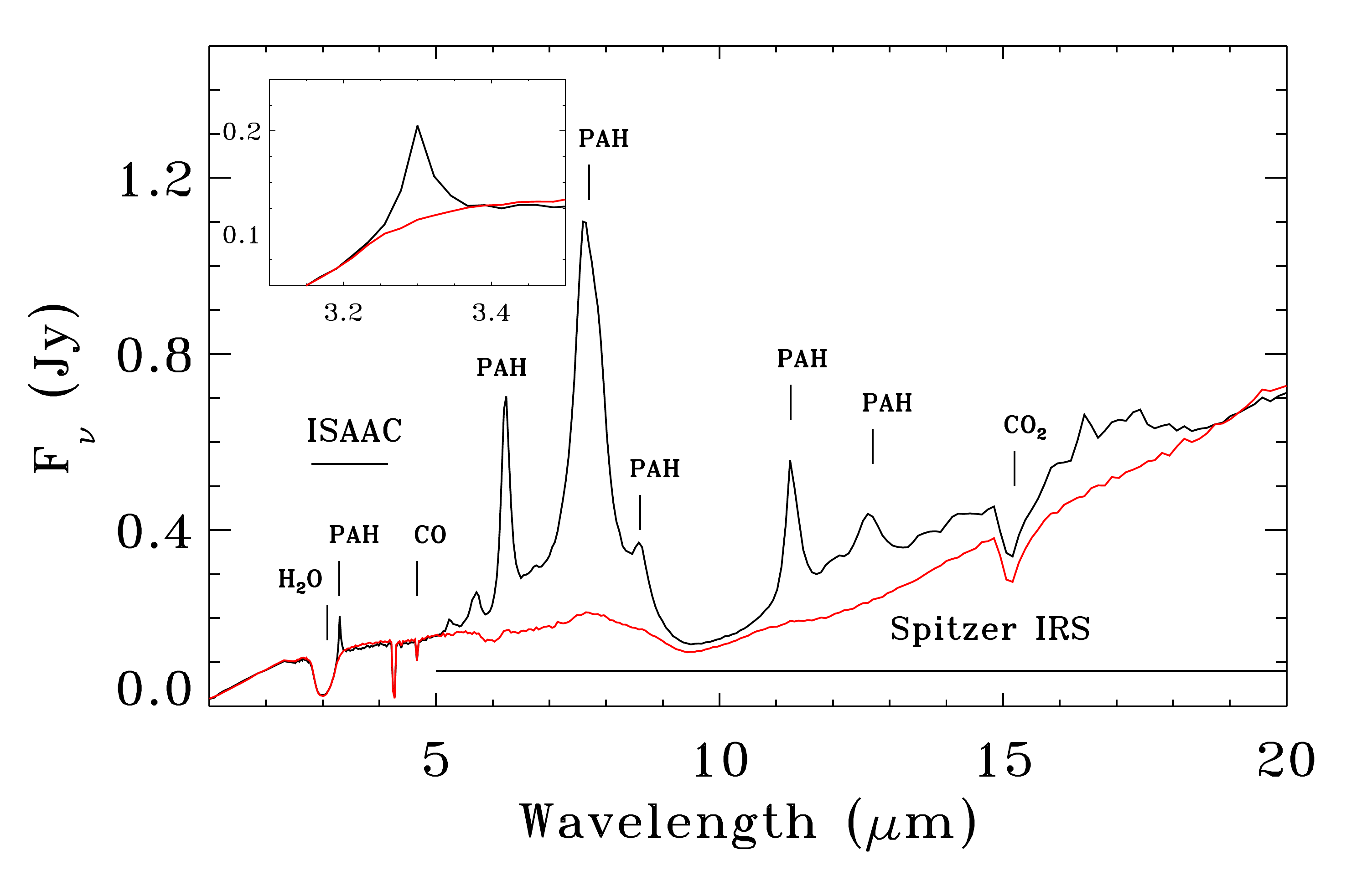}
  \caption{Model spectra at $i=45\degr$ for template parameters listed in Table \ref{embpah:tab:modelparams}, with PAHs (black) and without PAHs (red). Major PAH and absorption features are indicated, as well as the wavelength coverage of the presented ISAAC and Spitzer IRS observations. The inset shows a blow-up of the 3.3 \micron\ feature on the red wing of the water absorption band.}
  \label{embpah:fig:modelspec}
\end{figure}

A spectrum of the template model, with and without PAHs, is shown in Figure\ \ref{embpah:fig:modelspec}. The main features in the spectrum are the PAH features at 3.3, 6.2, 7.7, 8.6, 11.2 and 12.7 \micron, as well as water, CO, silicate and CO$_2$ ice absorption features at 3.08, 4.67, 9.7 and 15.2 \micron\ respectively. The wavelength coverage of the ISAAC L-band and Spitzer IRS spectroscopy is indicated.

\subsection{Modeling results}
The apparent absence of PAH features toward the majority of low-mass embedded class I sources could have a number of explanations. 

First, PAH molecules are primarily excited by UV photons. The precise shape and strength of the radiation field inside an embedded object is not well-known but in the class I sources the central source has already formed and is the main energy source inside the envelope. The presence or absence of excess UV will influence the PAH emission features. 
In particular, the high opacity of the envelope at UV and optical wavelengths, which provide the main excitation of PAHs, will constrain the region where PAHs can be excited to a small radius. 

Second, even if the infrared PAH features are produced in the disk or inner envelope close to the star, the dust in the surrounding envelope may provide too high extinction in the optical and mid-infrared for the (PAH) emission to escape, especially if the envelope mass is (still) relatively high compared to the disk. If an outflow cavity is present, the inclination at which the object is observed will affect the extinction by the envelope.

Third, the abundance of small PAH molecules in the gas phase may be significantly lower due to freeze-out due to the low temperatures and high densities in the interior of the molecular core. 

In practice, a combination of the above three causes will apply simultaneously. To estimate their effects, models are run, varying several parameters including radiation field, PAH abundance and the mass of the envelope.

\subsubsection{Luminosity of the central source and presence of UV excess}
\begin{figure}
  \centering
  \includegraphics[width=\columnwidth]{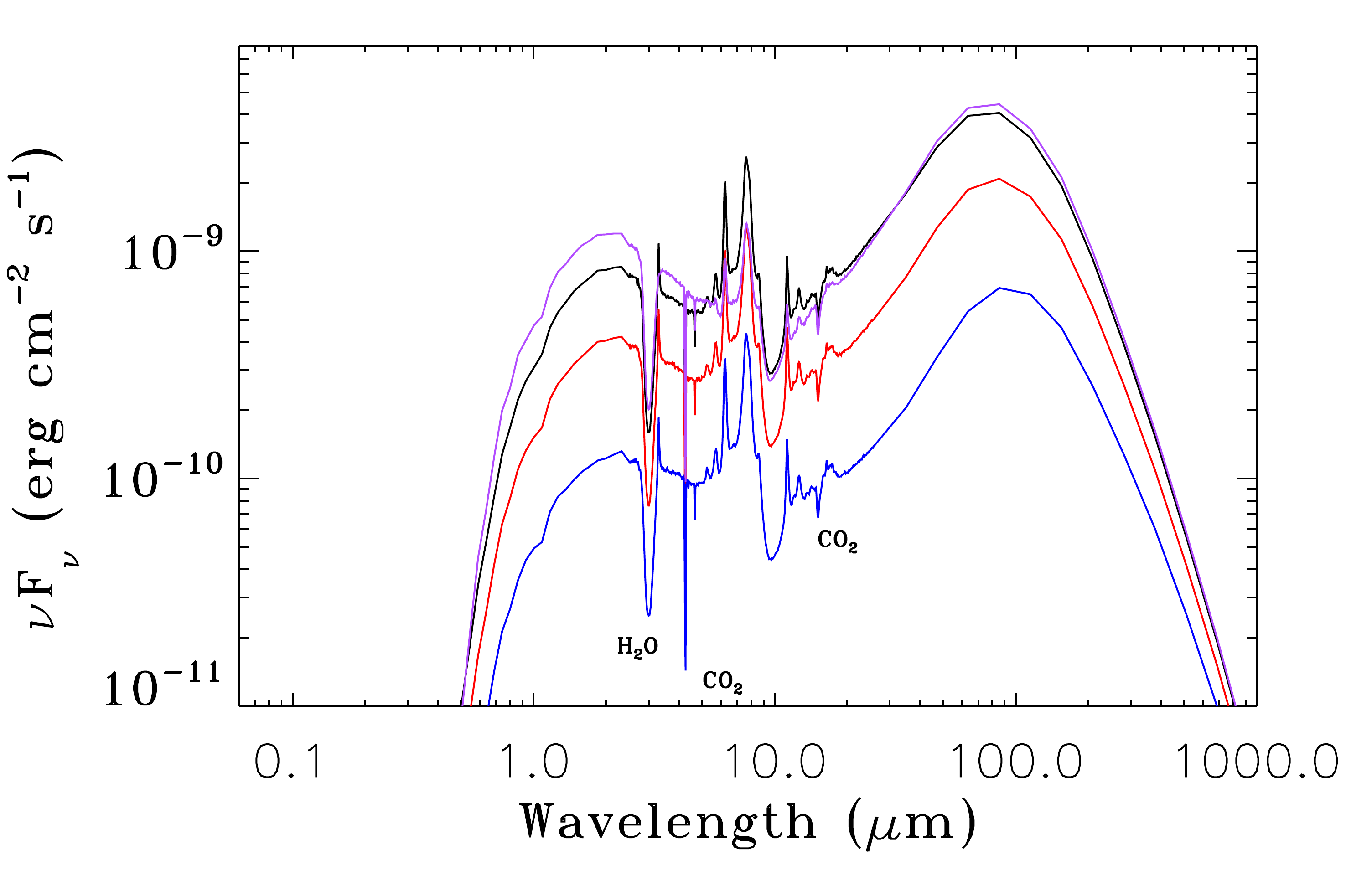}
  \includegraphics[width=\columnwidth]{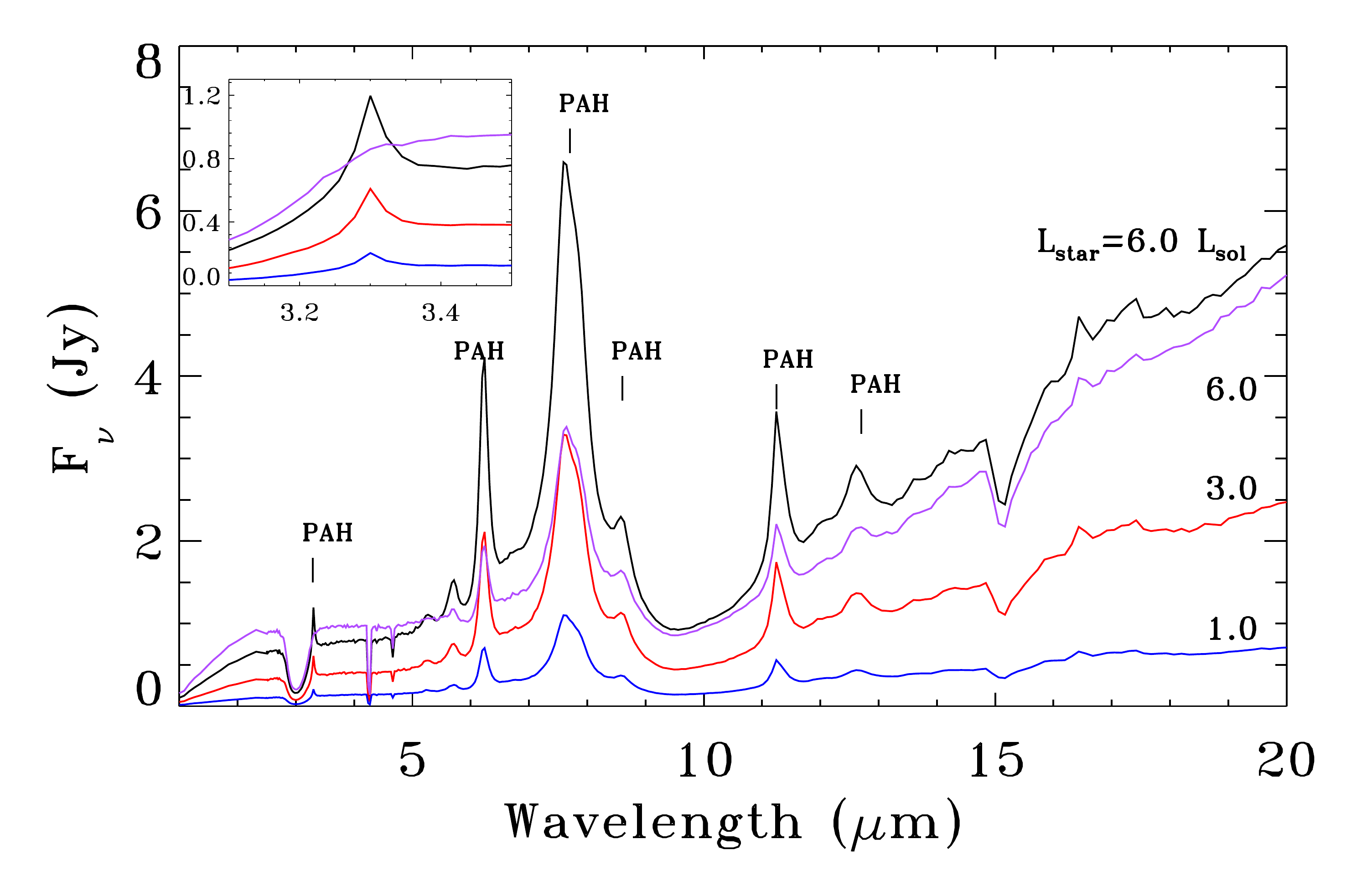}
  \caption{Model SEDs (top) and blow-up of spectra (bottom) at $i=45\degr$ for template model parameters and $L_*$ varying between 6, 3, and 1 L$_{\odot}$ (black, red, blue respectively), all including UV excess. A model with $L_* = 6$ L$_{\odot}$ without UV excess is shown in purple. Major PAH and absorption features are indicated. The inset shows a blow-up of the 3.3 \micron\ feature on the red wing of the water absorption band.}
  \label{embpah:fig:uvexcess}
\end{figure}

Models with total stellar luminosity $L_*$ varying between 1, 3 and 6 L$_{\odot}$ are shown in Fig.\ \ref{embpah:fig:uvexcess}. In addition, a model without the Draine field for UV excess is shown for $L_* = 6$ L$_{\odot}$. 
Increasing the luminosity by a factor of 6 increases the line flux of the PAH features by the same amount while the PAH feature / continuum ratio is reduced by at most 20\%. Excluding the excess UV field, while preserving the total stellar luminosity, decreases the PAH feature to continuum ratios by $\sim 3$. Interestingly, the 3.3 \micron\ feature is affected most because it requires higher energies to be excited than the other features. For typical luminosities observed toward low-mass embedded protostars, PAH features should be detectable if they have ISM abundances, even if no UV excess is present.

%############
\subsubsection{Mass of the envelope}
\begin{figure}
  \centering
  \includegraphics[width=\columnwidth]{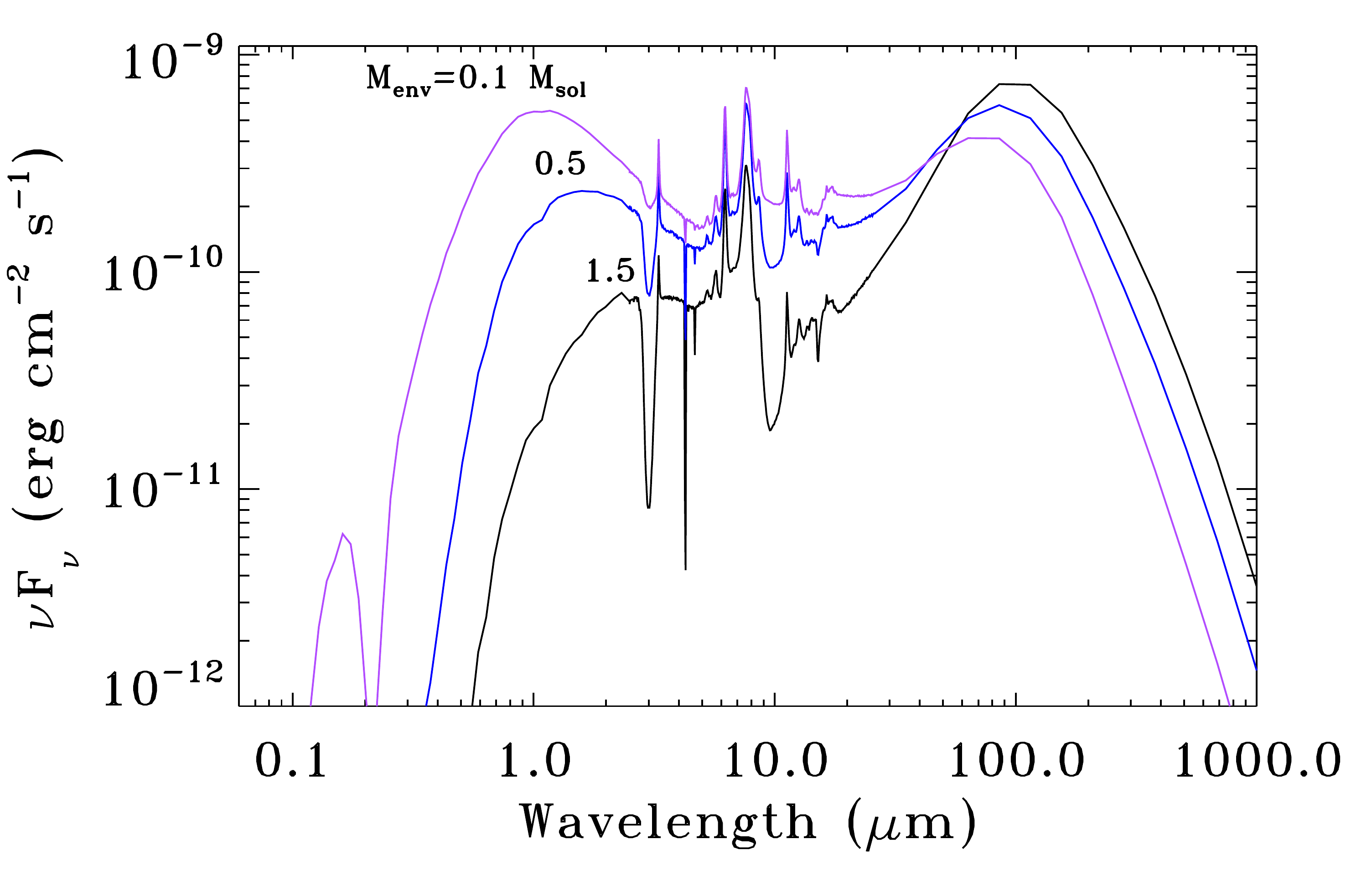}
  \includegraphics[width=\columnwidth]{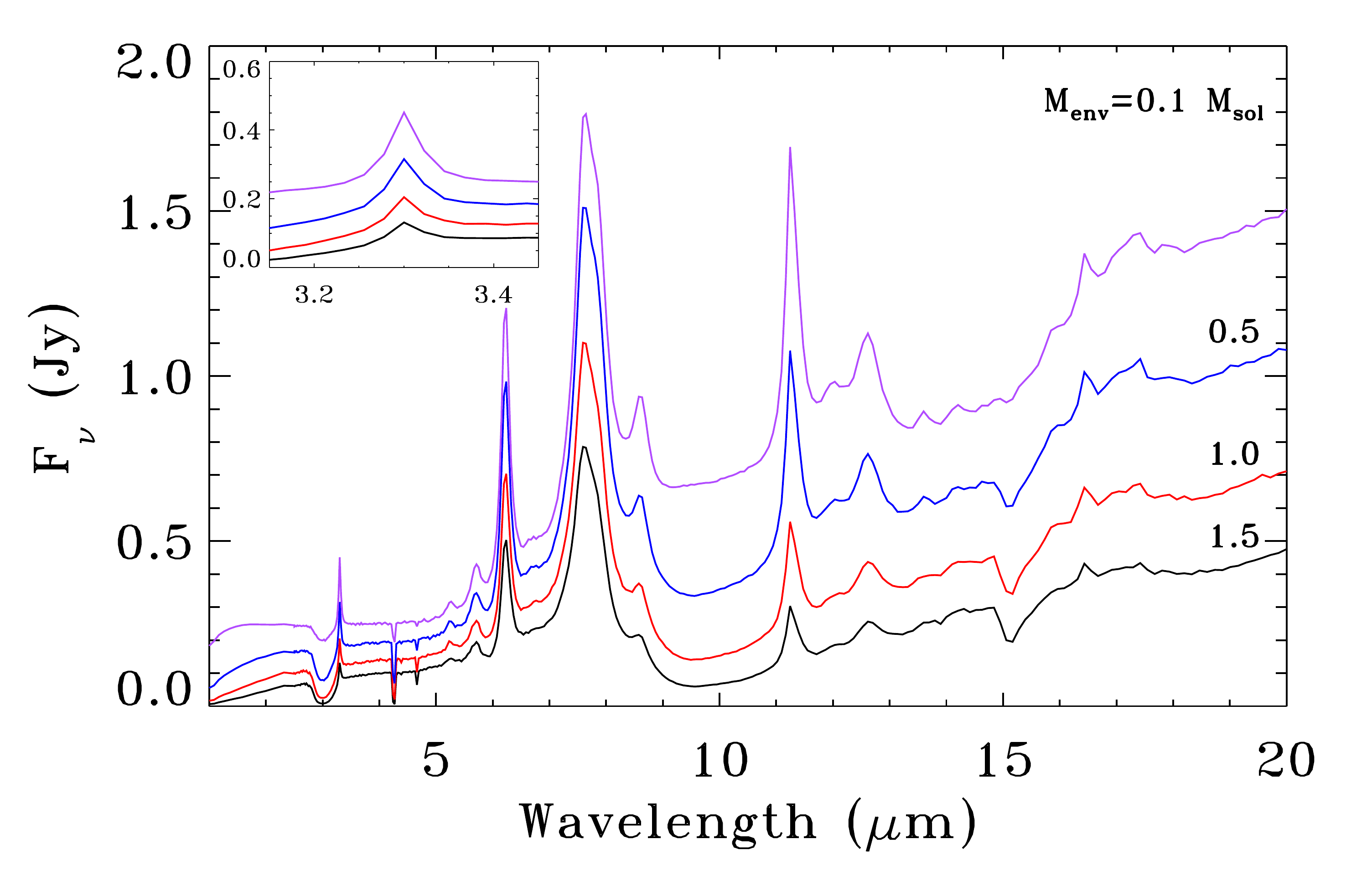}
  \caption{Model SEDs (top) and blow-up of spectra (bottom) with $M_{\mathrm{env}} = 1.5$ (black), 1.0 (red), 0.5 (blue) and 0.1 (purple) $M_{\odot}$ at $i = 45$\degr. The curve for 1.0 $M_{\odot}$ was omitted from the SED plot for clarity. The inset shows a blow-up of the 3.3 \micron\ PAH feature.}
  \label{embpah:fig:model_menv}
\end{figure}
Models with the mass of the envelope $M_{\mathrm{env}}$ varying from 0.1, 0.5, 1.0 to 1.5 $M_{\odot}$ are shown in Fig.\ \ref{embpah:fig:model_menv}, at $i = 45$\degr.
In all models the PAH features are clearly present. Increasing the envelope mass by a factor of 15 results in a decrease of the emission of the central source while the sub-mm emission and the strength of the absorption features increase. The PAH features decrease in peak flux by a factor of 3-4, but are in no case extinguished by the continuous extinction or silicate absorption features. This is consistent with the conclusions of \citet{man99} for higher mass YSO's.

\begin{figure}
  \centering
  \includegraphics[width=\columnwidth]{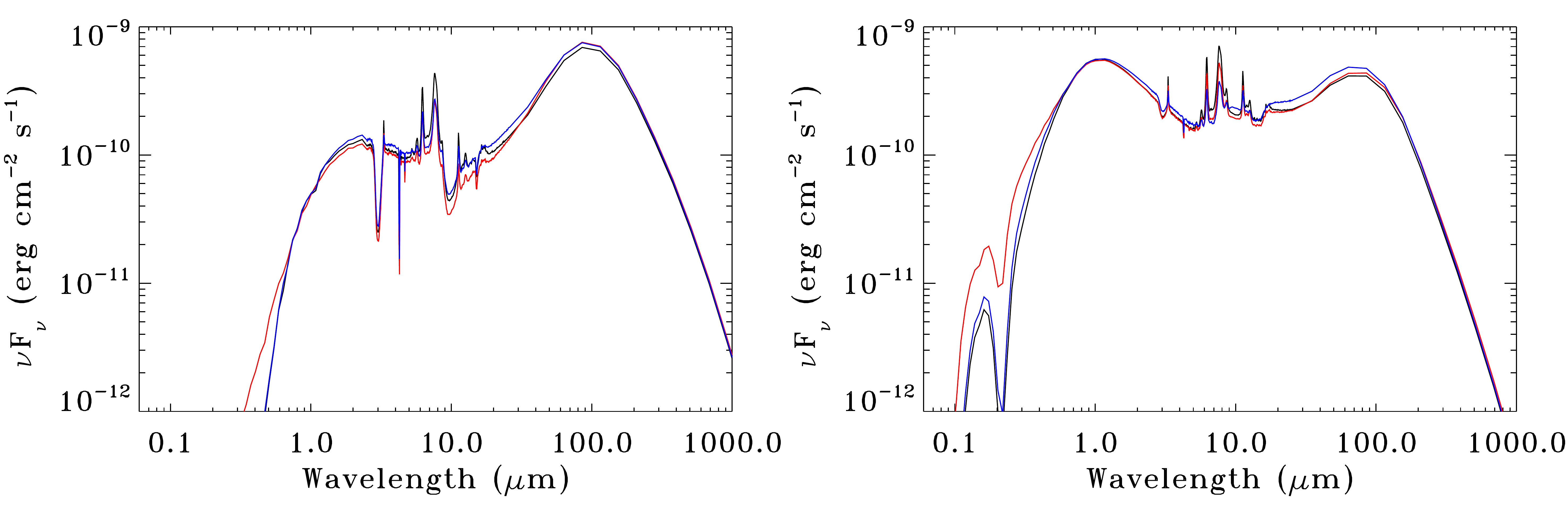}
  \includegraphics[width=\columnwidth]{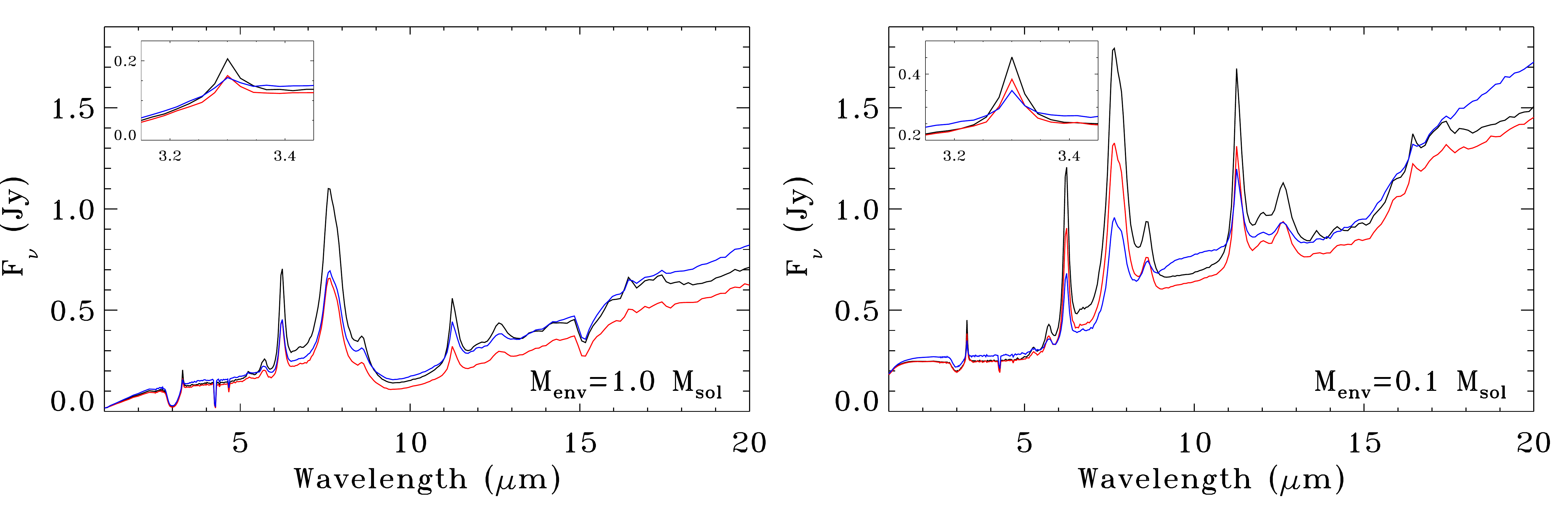}
  \caption{Model SEDs (top) and blow-up of spectra (bottom) with PAHs in both envelope and disk (black), only in the disk (red) and only in the envelope (blue), at $i = 45$\degr, for $M_{\mathrm{env}}$ = 1.0 (left) and 0.1 $M_{\odot}$ (right). The inset shows a blow-up of the 3.3 \micron\ PAH feature.}
  \label{embpah:fig:model_pahloc}
\end{figure}
In the template models, PAHs are located in both the envelope and the disk. To test the influence of the envelope further, model setups including PAHs only in the disk or only in the envelope were performed. A comparison is shown in Fig.\ \ref{embpah:fig:model_pahloc}, for $M_{\mathrm{env}}$ = 1.0 and 0.1 $M_{\odot}$.
Including PAHs only in the disk results in a decrease of peak flux of the PAH features by a factor $\sim$2. In the SED, the absence of PAHs in the envelope leads to less absorption of UV emission, which appears stronger. The mass of the envelope has no significant effect on the strength of the PAH features, although the underlying continuum changes.
Including PAHs only in the envelope also results in a typical decrease of the PAH feature peak flux, by about a factor of 1.5. Even if the PAHs are located only in the envelope, PAHs should be detectable if they are present at ISM abundance. 

%############
\subsubsection{Outflow cavity and inclination}
\begin{figure}
  \centering
  \includegraphics[width=\columnwidth]{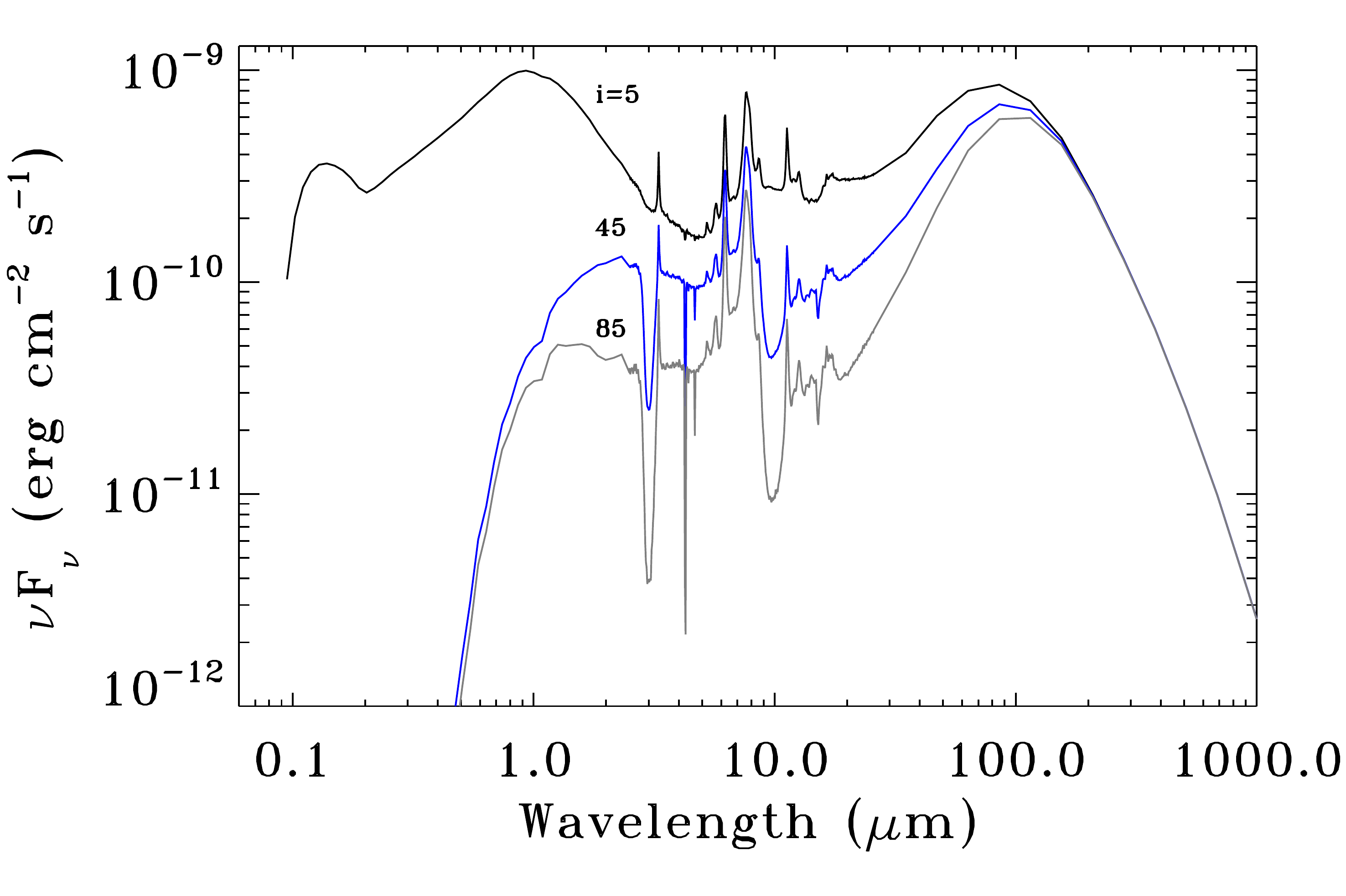}
  \includegraphics[width=\columnwidth]{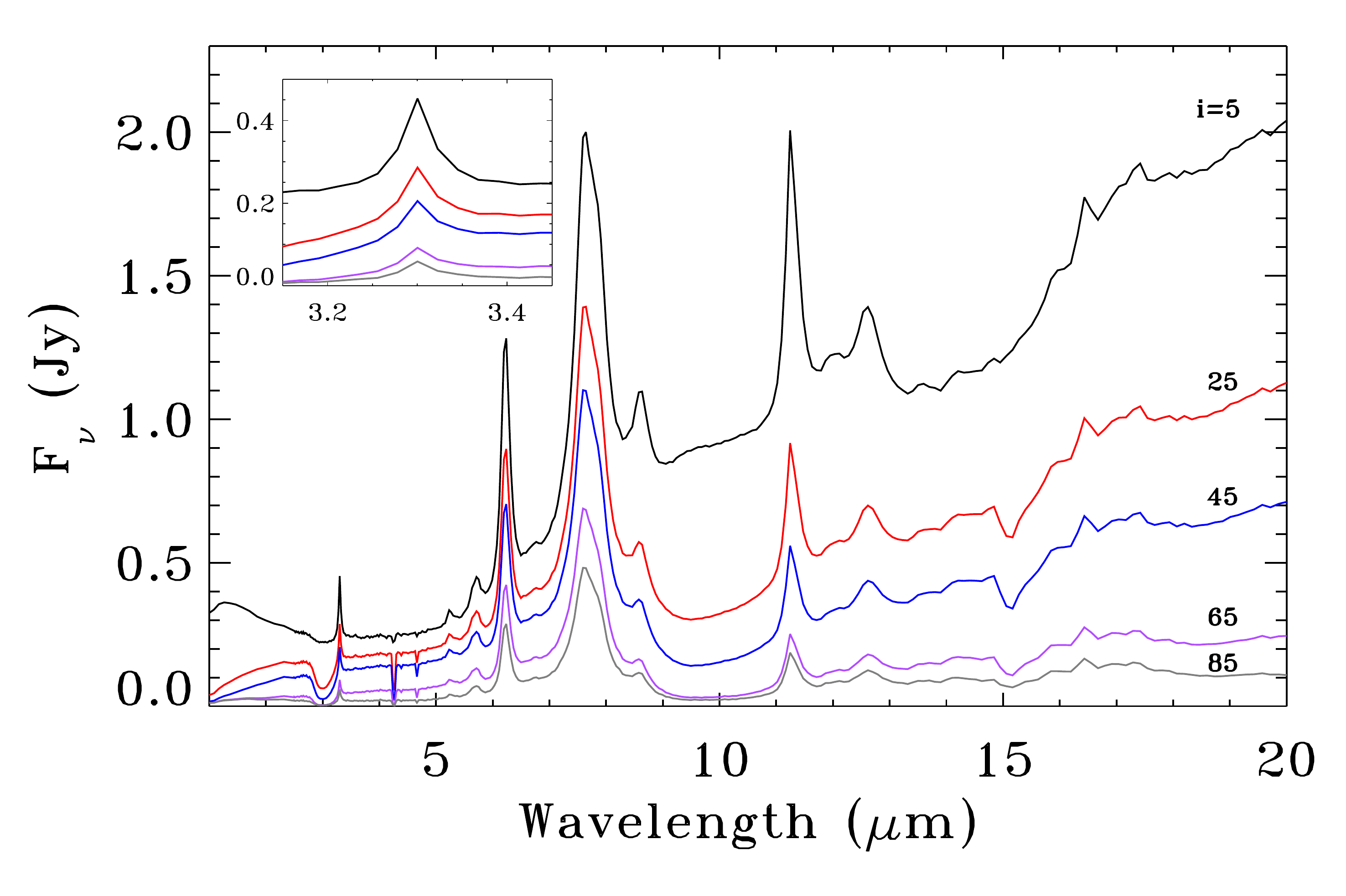}
  \caption{Model SEDs (top) and blow-up of spectra (bottom) for $i = 5$, 25, 45, 65 and 85\degr (black, red, blue, purple, gray, respectively). Curves for $i=25$ and 65\degr\ were omitted from the SED plot for clarity. The inset shows a blow-up of the 3.3 \micron\ PAH feature.}
  \label{embpah:fig:model_incl}
\end{figure}
Embedded class I objects are known to have outflows \citep[e.g.,\ ][]{hog98,arc06} so that an outflow cavity is included in our model envelope. Because of the presence of the outflow cavity, the inclination at which the object is observed is important, because at near-pole-on orientation one observes directly the central source and the disk. At larger inclinations, the envelope will be obscuring the disk.

A template model with the template PAH abundance and an envelope mass of 1.0 $M_{\odot}$, seen at varying inclination angles between 5 and 85\degr\ is shown in Fig.\ \ref{embpah:fig:model_incl}.
At an inclination of $i=5$\degr\ the stellar radiation field (blackbody + scaled Draine field) is directly visible and no absorption features are present. Between inclination of 5\degr\ and 25\degr, i.e., down the cavity and through the envelope, the appearance of the SED and the PAH features changes rapidly \citep[e.g.,][]{whi03,cra08}. The emission of the central star becomes obscured and the strength of the PAH features decreases by about a factor of 2. At increasing inclination the PAH features become weaker, but still dominate the spectrum. Ice absorption features can be seen at 3, 4.2, 6 and 15 \micron. 
At 85\degr, the disk is observed almost edge-on and the PAH emitting regions are largely masked by the disk itself. The remaining features in the spectrum extracted for an infinitely large aperture are due to scattered emission, originating from higher up in the disk atmosphere. Observed PAH feature strengths will depend on the pointing and orientation of the limited aperture slit on the embedded source.

%############
\subsubsection{PAH abundance}
\begin{figure}
  \centering
  \includegraphics[width=\columnwidth]{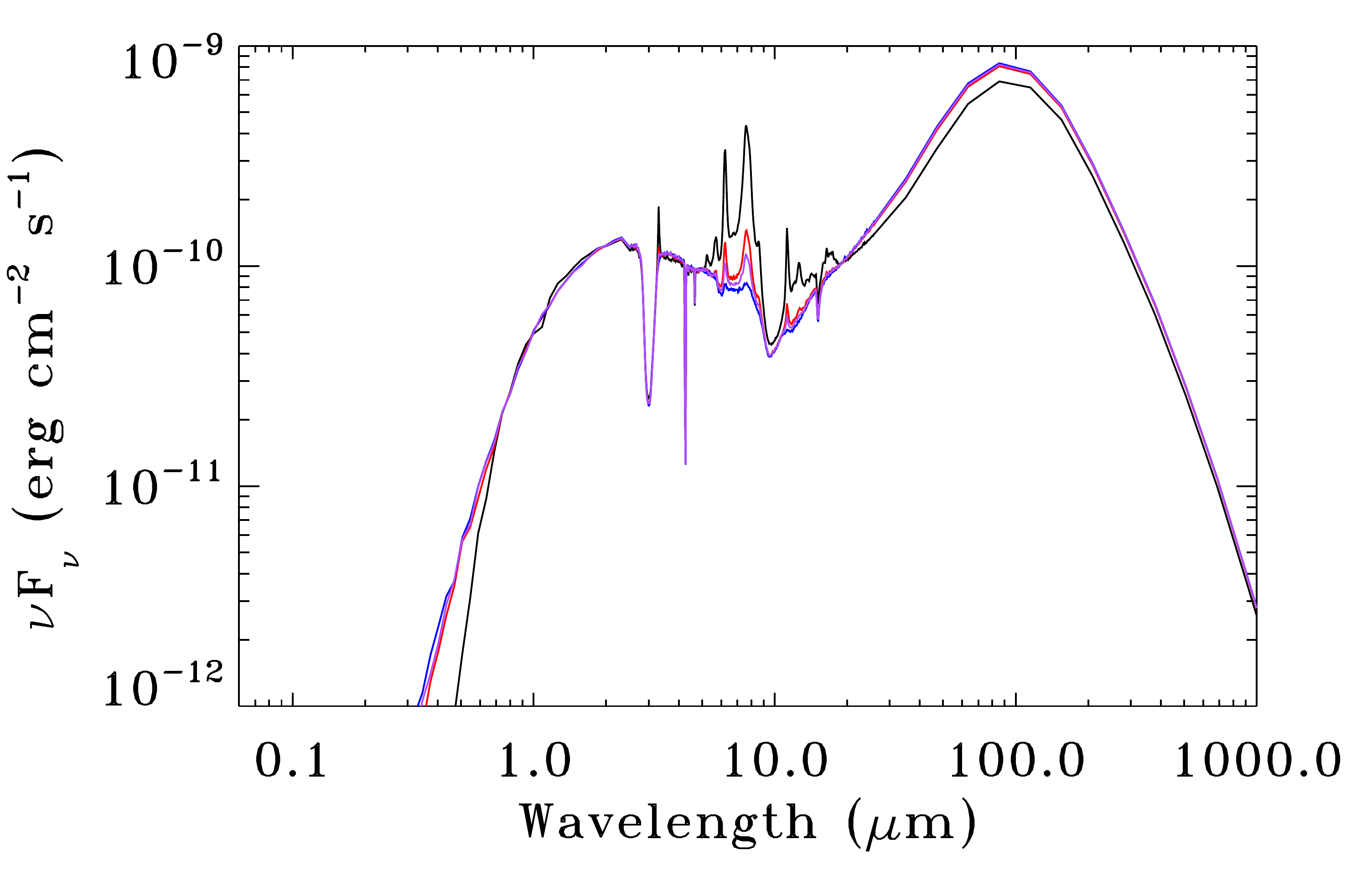}
  \includegraphics[width=\columnwidth]{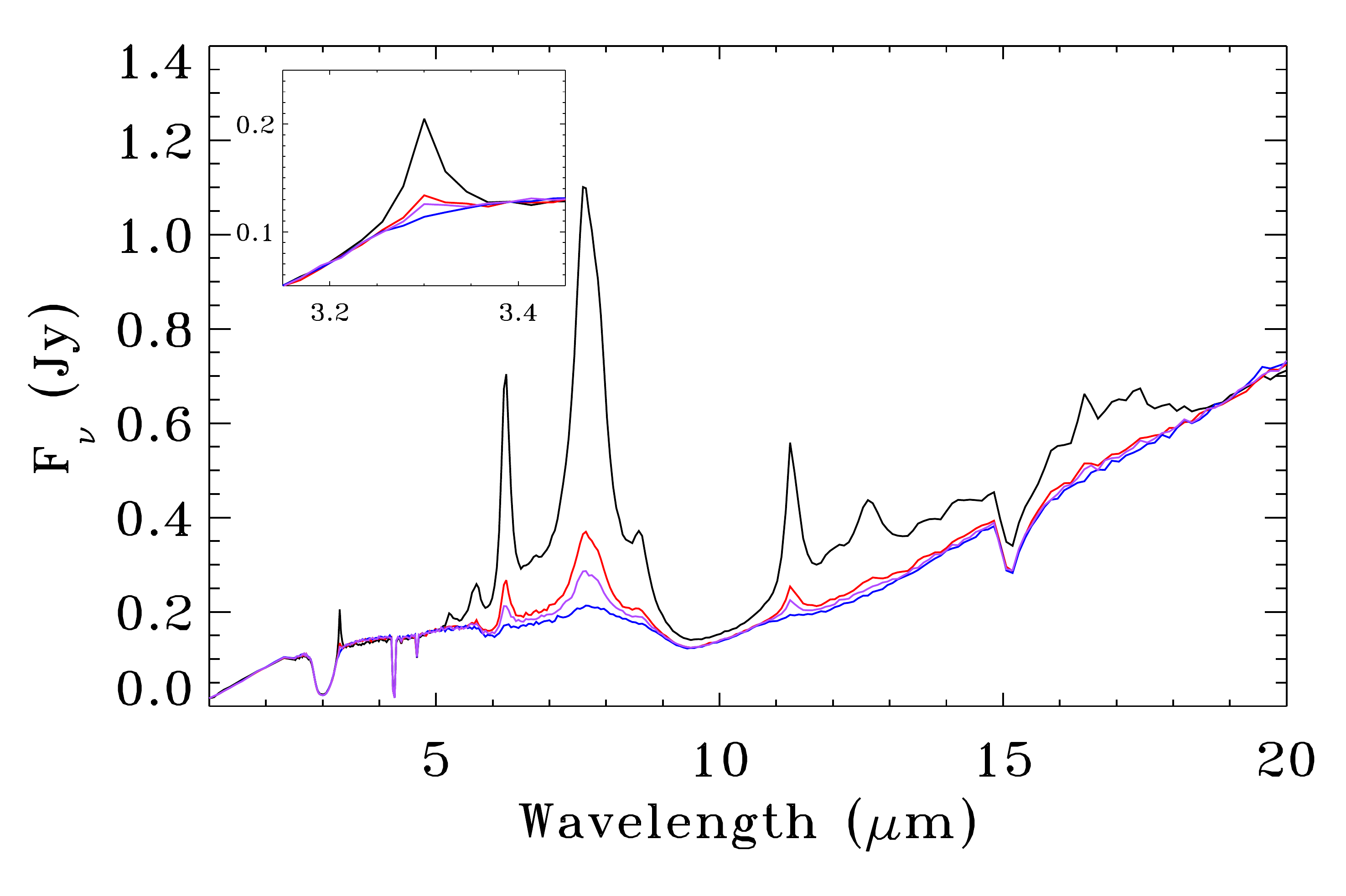}
  \caption{Model SEDs (top) and blow-up of spectra (bottom) with template PAH abundance (black), and factors of 10 (red), 20 (purple) and 100 lower (blue) at $i = 45$\degr. The inset shows a blow-up of the 3.3 \micron\ PAH feature.}
  \label{embpah:fig:model_pahabun}
\end{figure}
Models with the total PAH abundance (both ionized and neutral species) varying between the template abundance, and factors of 10, 20 and 100 lower, are shown in Fig.\ \ref{embpah:fig:model_pahabun}. The envelope mass in these models is 1.0 $M_{\odot}$.
Reducing the PAH abundance by a factor of 10 has the straightforward result of reducing the PAH peak flux by a factor of 3. The 3.3 \micron\ feature is notably weak in all models, and already at a factor of 10 lower abundance not anymore detectable. Decreasing the PAH abundance also leads to an increase in the UV emission and continuum emission at far-IR wavelengths. The strongest feature, 7.7 \micron, is visible down to an abundance of $1 \times 10^{-8}$ relative to H, i.e., a value 50x lower than the ISM.

\subsection{Summary and caveats}
We find in all models that the 7.7 \micron\ feature is noticeably the strongest PAH feature, with the highest feature to continuum ratio. This is in sharp contrast with observational and modeling results for non-embedded class II stars with disks, where the silicate emission feature largely obscures the contribution of the 7.7 and 8.6 \micron\ features.  There is not any readily visible 7.7 \micron\ PAH feature in our observations. Increasing the envelope mass and decreasing PAH abundance, luminosity or UV excess all lead to weaker PAH features. The inclination has the largest influence on the PAH feature strength in the range of inclinations where the line-of-sight passes from through the envelope to through the outflow cavity. Strong PAH features are predicted to occur when observing down the outflow cavity. This situation can be recognized by a strong UV and optical part of the SED, as well as a lack of absorption features. For typical values for the luminosity, UV excess, inclination and envelope mass, derived from observational studies of embedded protostars, and assuming an ISM abundance, the PAH features should be detectable. The low detection rate of PAHs toward these sources thus suggests that the abundance of PAHs is at least 20--50x lower than in the ISM. Decreasing luminosity, UV excess or envelope mass compared with the template model changes this conclusion to a typical factor of 10--20, compared with the ISM abundance.

One caveat remains with respect to the effect of scattering. The radiative transfer code used here assumes all scattering to be isotropic. This does not allow us to treat the effect of preferentially forward scattering of UV radiation through the outflow cavity. In the current implementation, our models could overestimate the amount of radiation received by the disk as well as overestimate the strength of the PAH features. Thus, our inferred abundance limits are on the low side. As a second caveat, an external interstellar radiation field is not included in these models. Interstellar UV radiation from outside the embedded object may become a significant source of excitation for PAHs in the envelope at large radii and would increase the predicted PAH feature strength and thus constrain the PAH abundance upper limits to smaller values, going in opposite directions.

\subsection{PAH evolution from clouds to disks}
In both class I and II low-mass objects that we have studied, we find very low detection rates for PAHs.
Comparison with model predictions suggest that for typical conditions in class I protostars, the absence of exciting radiation and attenuation by foreground material can be excluded as reason for the absence of the PAH features. In addition, the predicted feature / continuum ratio is well within the achieved sensitivity of the observations. This thus implies an absence of carrier as the reason the for absence of  features, and the PAH abundance in the embedded protostar phase is estimated here to be at least a factor of 10--20 times lower than for typical (number of carbon atoms $N_C = 100$) PAHs in the ISM. This lower abundance is in the same range as those inferred toward T Tauri stars in \citet{gee06}.  Thus, even though the detection rate in the Class I objects is even lower than that found toward disks, the inferred abundances are comparable. Note that the few Class II disks in which we detect PAHs are mostly G-stars. The disk sample in Geers et al.\ (2006) is more biased toward higher mass stars.

What would cause this lower abundance in the embedded phase, and is this mechanism the same in the class II phase? In a cold dense environment, two possibilities for lowering the abundance of a species are recognized: coagulation or dust growth and freeze-out onto larger grains. 

The freeze-out of PAHs onto grains could explain the lower PAH abundance in these objects. PAH features in absorption have been observed in the ISM, and laboratory spectra of PAHs condensed into H$_2$O rich ices suggest that PAH absorption features should also be observable in astrophysical ices, such as in dense clouds \citep[][and references therein]{ber07}. During this phase, if exposed to UV radiation, PAHs in water ice can become significantly ionized, thus promoting ion-mediated reactions in these ices with implications for astrochemistry \citep{ehr06,gud06}. In the class II phase, the direct irradiation of the disk by the central star can increase the temperature in the surface layers of the disk, evaporating the ice and depositing the enclosed PAHs back into the gas, thereby increasing the abundance of the PAHs again. Recent observations of edge-on disks show evidence for the continued presence of a reservoir of ice in the interior of disks \citep{pon05,ter07} which can be brought to the surface by vertical mixing. If the PAHs in the ices react to form larger carbonaceous materials, then only a fraction of PAHs might come back off the grain when the ice layer evaporates. This could be a possible explanation for the low PAH abundance inferred for the majority of T Tauri disk surface layers.
In a study of the ice features of the same c2d sample of embedded sources, \citep{boo08} show spectra over the 5--10 \micron range where absorptions due to PAHs in ices occur. No obvious features are detected, but the laboratory spectroscopy is not well known and some PAH bands could be hidden in the 6 \micron\ complex, especially on the long wavelength side of the C2 component in \citet{boo08}. Assuming a typical (but uncertain) oscillator strength for PAHs, up to 10\% of the carbon could be locked up in PAHs in the ices (Boogert, priv.\ comm.). These limits are consistent with the amount of freeze-out found in this work.

PAHs growing and/or agglomerating to typically 100x larger particles could be an alternative explanation of lowering the PAH abundance. This process could take place in the high density regions of the disk and envelope, in both the class I and II phase, and does not exclude freeze-out as an option. Observationally, the presence of broad emission plateaus between 6--9, 11--13 and 15--20 \micron\ has been attributed to large 200--2000 carbon atom PAHs  and PAH clusters \citep{all89,ker00}, which will lose their characteristic PAH emission features, thus additionally lowering the feature / continuum of these features. The detection of individual features toward T Tauri disks shows that this process is not necessarily dominant, and that either small PAHs are still released from the ice in this phase and/or that small PAHs are still being created from, e.g., the destruction of larger grains or agglomerates through collisions. 

%############
\section{Conclusions}
No PAHs are detected toward any of the 53 confirmed embedded sources in neither our Spitzer nor ISAAC surveys. PAHs are detected at 3.3 \micron\ towards 1 of the 17 confirmed disk sources, IRS~48, and at 11.2 \micron\ for 1 of the 10 sources with an uncertain or borderline embedded classification, GY~23. 
For all 12 sources with ISAAC and Spitzer spectra, no PAH features are detected in either.  In total, PAH features are detected toward at most 1 out of 63 (candidate) embedded protostars (1.6\%), even lower than observed for class II T Tauri stars with disks (11--14\%). 

The low PAH detection rate is compared with radiative transfer model calculations. It is shown that the effects of envelope mass on the absorption of PAH features is not enough to obscure PAH features. The 7.7 \micron\ feature is predicted by models as the best tracer of PAH emission, while the 3.3 \micron\ feature is relatively weak. 

Assuming typical class I stellar and envelope parameters, the absence of PAH emission is most likely explained by the absence of emitting carriers through a much lower PAH abundance in the gas, e.g., due to freeze-out of PAHs on icy layers on dust grains or agglomeration. Our inferred low abundances are similar to those found for disks around T Tauri stars. Thus PAHs are expected to be removed from the gas already at earlier stages of star- and planet formation and enter the disks frozen out on grains. Further searches for PAH absorption features are needed to constrain the presence of PAHs in icy grains.

%#######################################################
\begin{acknowledgements}
%#######################################################
The authors thank R. Visser for his models and comments.
Support for this work, part of the Spitzer Legacy Science Program, was provided by NASA through contracts 1224608, 1230779 and 1256316 issued by the Jet Propulsion Laboratory, California Institute of Technology,
under NASA contract 1407. A.C. was supported by a fellowship from the European Research Training Network ``The Origin of Planetary Systems'' (PLANETS, contract number HPRN-CT-2002-00308) at Leiden Observatory and by a Marie Curie Intra-European Fellowship from the European Community (contract number FP6-024227) at Observatorio Astron\'omico Nacional. Astrochemistry in Leiden is supported by a NWO Spinoza grant and a NOVA grant.
\end{acknowledgements}

\clearpage
\appendix
\section{Tables, figures}
{\scriptsize
\longtab{2}{
\begin{longtable}{llllll}

\caption[Summary of observations.]{Summary of observations. Line flux in W m$^{-2}$.} 
\label{embeddedpah:tab:obssum} \\

\hline
\hline
Target          & RA       & Dec        & class      & Line flux   & Comments\\
                & [J2000]  & [J2000]    &            & 3.3 \micron &         \\
\hline
\endfirsthead

\hline
\hline
Target          & RA & Dec &  class & Line flux    & Comments\\
                & [J2000]  & [J2000]  &            & 3.3 \micron &         \\
\hline
\endhead

\hline
\endfoot

\hline
\endlastfoot

Perseus\\
\hline
LDN1448 IRS1            & \ra{3}{25}{09}{4}  & \dec{+30}{46}{21}{7} & embedded     & -         & Spitzer \\ 
LDN1448 NA              & \ra{3}{25}{36}{5}  & \dec{+30}{45}{21}{2} & embedded     & - 	    & Spitzer \\ 
IRAS 03245+3002         & \ra{3}{27}{39}{0}  & \dec{+30}{12}{59}{4} & embedded     & - 	    & Spitzer \\ 
L1455 SMM1              & \ra{3}{27}{43}{2}  & \dec{+30}{12}{28}{8} & embedded     & - 	    & Spitzer \\ 
L1455 IRS3              & \ra{3}{28}{00}{4}  & \dec{+30}{08}{01}{3} & embedded     & - 	    & Spitzer \\ 
IRAS 03254+3050         & \ra{3}{28}{34}{5}  & \dec{+31}{00}{51}{1} & embedded     & - 	    & Spitzer \\ 
IRAS 03271+3013         & \ra{3}{30}{15}{2}  & \dec{+30}{23}{48}{8} & embedded     & - 	    & Spitzer \\ 
IRAS 03301+3111         & \ra{3}{33}{12}{8}  & \dec{+31}{21}{24}{1} & embedded     & - 	    & Spitzer \\ 
B1-a                    & \ra{3}{33}{16}{7}  & \dec{+31}{07}{55}{2} & embedded     & - 	    & Spitzer \\ 
B1-c                    & \ra{3}{33}{17}{9}  & \dec{+31}{09}{31}{0} & embedded     & - 	    & Spitzer \\ 
SSTc2d J033327.3+310710 & \ra{3}{33}{27}{3}  & \dec{+31}{07}{10}{2} & embedded     & - 	    & Spitzer \\ 
HH 211-mm               & \ra{3}{43}{56}{8}  & \dec{+32}{00}{50}{4} & embedded     & - 	    & Spitzer \\
IRAS~03439+3233         & \ra{3}{47}{05}{4}  & \dec{+32}{43}{08}{4} & embedded/borderline     & - 	    & Spitzer \\ 
IRAS~03445+3242         & \ra{3}{47}{41}{6}  & \dec{+32}{51}{43}{9} & embedded     & - 	    & Spitzer \\ 
\hline
Taurus \\
\hline
LDN~1489~IRS            & \ra{04}{04}{42}{9} & \dec{+26}{18}{56}{3} & embedded  & $\leq$8.8E-16 & ISAAC \\ 
\hline
Orion \\
\hline
Reipurth 50             & \ra{05}{40}{27}{7} & \dec{-07}{27}{32}{1} &  embedded   & $\leq$1.0E-15 & ISAAC \\ 
TPSC~78                 & \ra{05}{35}{14}{1} & \dec{-05}{23}{38}{4} &  uncertain   & $\leq$2.3E-16 & ISAAC \\
TPSC~1                  & \ra{05}{35}{14}{5} & \dec{-05}{23}{54}{7} &  uncertain   & $\leq$1.0E-16 & ISAAC \\ 
\hline
Vela \\
\hline
HH~46 / IRAS~08242-5050 & \ra{08}{25}{43}{8} & \dec{-51}{00}{35}{6} & embedded  & $\leq$1.2E-16 & ISAAC, Spitzer\\ 
IRAS~08261-5100         & \ra{08}{27}{38}{9} & \dec{-51}{10}{37}{2} & embedded  & -             & Spitzer \\
LLN~20                  & \ra{08}{47}{39}{4} & \dec{-43}{06}{08}{-} & embedded & $\leq$1.1E-16 & ISAAC \\ 
LLN~33                  & \ra{08}{57}{36}{8} & \dec{-43}{14}{35}{-} & embedded & $\leq$1.3E-16 & ISAAC \\ 
LLN~47                  & \ra{09}{09}{25}{6} & \dec{-45}{22}{51}{-} & embedded & $\leq$5.3E-16 & ISAAC \\ 
\hline
Chamaeleon \\
\hline
Ced 110 IRS4        & \ra{11}{06}{46}{6} & \dec{-77}{22}{32}{5} & embedded  &               & Spitzer \\
IRAS~11068-7717         & \ra{11}{08}{15}{1} & \dec{-77}{33}{53}{2} & uncertain  & $\leq$5.5E-15 & ISAAC\\ 
Cha~IRN                 & \ra{11}{08}{38}{2} & \dec{-77}{43}{51}{7} & embedded  & $\leq$4.2E-16 & ISAAC \\ 
Cha~INa~2               & \ra{11}{09}{36}{6} & \dec{-76}{33}{39}{-} & disk  & $\leq$5.1E-16 & ISAAC \\ 
IRAS 12553-7651         & \ra{12}{59}{06}{6} & \dec{-77}{07}{40}{1} & embedded  &               & Spitzer \\ 
\hline
Ophiuchus \\
\hline
VSSG1               & \ra{16}{26}{18}{8} & \dec{-24}{28}{49}{5} & disk & $\leq$9.5E-16 & ISAAC, Spitzer\\
GSS30~IRS~1         & \ra{16}{26}{21}{4} & \dec{-24}{23}{04}{2} & embedded    & $\leq$8.6E-16 & ISAAC, Spitzer\\ 
GY~23                 & \ra{16}{26}{24}{1} & \dec{-24}{24}{48}{2} & embedded/borderline   & -		   & Spitzer \\
VLA~1623-2418            & \ra{16}{26}{26}{4} & \dec{-24}{24}{30}{2} & embedded    & -		   & Spitzer \\
IRS~14                  & \ra{16}{26}{31}{0} & \dec{-24}{31}{05}{2} & disk    & -		   & Spitzer \\
WL~12                   & \ra{16}{26}{44}{2} & \dec{-24}{34}{48}{4} & embedded    & $\leq$2.6E-16 & ISAAC, Spitzer \\
OphE-MM3                & \ra{16}{27}{05}{9} & \dec{-24}{37}{08}{0} & disk    & -		   & Spitzer\\
GY~224                  & \ra{16}{27}{11}{2} & \dec{-24}{40}{46}{6} & disk & -		   & Spitzer\\
WL~19                   & \ra{16}{27}{11}{7} & \dec{-24}{38}{32}{3} & uncertain & -		   & Spitzer\\ 
WL~20S                  & \ra{16}{27}{15}{6} & \dec{-24}{38}{45}{6} & disk    & $\leq$2.0E-16 & ISAAC, Spitzer\\ 
WL~20E                  & \ra{16}{27}{15}{7} & \dec{-24}{38}{59}{8} & disk    & $\leq$2.1E-16 & ISAAC\\
IRS~37                  & \ra{16}{27}{17}{6} & \dec{-24}{28}{56}{6} & embedded    & -		   & Spitzer\\
IRS~42                  & \ra{16}{27}{21}{5} & \dec{-24}{41}{43}{1} & disk & $\leq$2.0E-15 & ISAAC \\
WL~6                    & \ra{16}{27}{21}{8} & \dec{-24}{29}{53}{2} & embedded    & -		   & Spitzer\\ 
CRBR~2422.8-3423    & \ra{16}{27}{24}{6} & \dec{-24}{41}{03}{1} & disk & $\leq$8.7E-16 & ISAAC, Spitzer\\
IRS~43                  & \ra{16}{27}{26}{9} & \dec{-24}{40}{50}{8} &  embedded   & $\leq$4.4E-15 & ISAAC \\
IRS~44                  & \ra{16}{27}{28}{0} & \dec{-24}{39}{33}{5} &  embedded   & $\leq$1.3E-15 & ISAAC \\ 
Elias~32                & \ra{16}{27}{28}{4} & \dec{-24}{27}{21}{2} & embedded & $\leq$8.4E-16 & ISAAC, Spitzer\\
IRS~46              & \ra{16}{27}{29}{4} & \dec{-24}{39}{16}{2} & disk    & $\leq$1.4E-15 & ISAAC, Spitzer\\ 
VSSG~17                 & \ra{16}{27}{30}{2} & \dec{-24}{27}{44}{3} & embedded    & $\leq$3.9E-16 & ISAAC, Spitzer\\
IRS~48              & \ra{16}{27}{37}{2} & \dec{-24}{30}{35}{0} &  disk   & 3.0E-15	   & ISAAC \\ 
IRS~51                  & \ra{16}{27}{39}{8} & \dec{-24}{43}{15}{1} & disk & $\leq$9.2E-16 & ISAAC \\ 
IRS~54                  & \ra{16}{27}{51}{8} & \dec{-24}{31}{45}{5} & disk   & $\leq$3.8E-16 & ISAAC \\
IRS~63                  & \ra{16}{31}{35}{7} & \dec{-24}{01}{29}{6} & embedded & $\leq$3.2E-15 & ISAAC, Spitzer \\
L1689-IRS5              & \ra{16}{31}{52}{1} & \dec{-24}{56}{15}{4} & uncertain    & -		   & Spitzer \\ 
IRAS 16293-2422B        & \ra{16}{32}{22}{6} & \dec{-24}{28}{32}{2} & embedded    & -		   & Spitzer \\
IRAS 16293-2422         & \ra{16}{32}{22}{9} & \dec{-24}{28}{36}{1} & embedded    & -		   & Spitzer \\ 
RNO~91                  & \ra{16}{34}{29}{3} & \dec{-15}{47}{01}{3} & embedded    & -		   & Spitzer \\ 
\hline							   	     
Serpens \\
\hline
SSTc2d J182901.8+02954  & \ra{18}{29}{01}{8} & \dec{+00}{29}{54}{2}  & uncertain    & - 	    & Spitzer \\
SSTc2d J182916.2+01822  & \ra{18}{29}{16}{2} & \dec{+00}{18}{22}{7}  & embedded    & - 	    & Spitzer \\
Serp-S68N               & \ra{18}{29}{48}{1} & \dec{+01}{16}{42}{6}  & embedded    & - 	    & Spitzer \\ 
EC69                    & \ra{18}{29}{54}{4} & \dec{+01}{15}{01}{8}  & uncertain    & - 	    & Spitzer \\ 
SVS~4-2                 & \ra{18}{29}{56}{6} & \dec{+01}{12}{59}{4}  & disk    & $\leq$6.4E-17 & ISAAC \\
Serp-SMM4               & \ra{18}{29}{56}{6} & \dec{+01}{13}{15}{2}  & embedded    & - 	    & Spitzer \\
EC~82                   & \ra{18}{29}{56}{9} & \dec{+01}{14}{46}{4}  & disk    & $\leq$4.9E-16 & ISAAC \\
EC~90A                  & \ra{18}{29}{57}{3} & \dec{+01}{14}{03}{7}  & embedded    & $\leq$2.1E-15 & ISAAC \\
EC~90B                  & \ra{18}{29}{57}{3} & \dec{+01}{14}{03}{7}  & embedded    & $\leq$7.4E-16 & ISAAC \\ 
EC88                    & \ra{18}{29}{57}{6} & \dec{+01}{13}{00}{5}  & embedded    & - 	    & Spitzer \\ 
SVS~4-3                 & \ra{18}{29}{56}{7} & \dec{+01}{12}{39}{0}  & disk    & $\leq$1.4E-16 & ISAAC \\
SVS 4-5                 & \ra{18}{29}{57}{6} & \dec{+01}{13}{00}{2}  & disk    & $\leq$1.6E-16 & ISAAC \\ 
SVS 4-9                 & \ra{18}{29}{58}{1} & \dec{+01}{12}{39}{4}  & disk    & $\leq$3.2E-16 & ISAAC \\
Serp-SMM3               & \ra{18}{29}{59}{2} & \dec{+01}{14}{00}{2}  & embedded    & - 	    & Spitzer \\
\hline
Coronis Australis \\
\hline
R~CrA~IRS~5A             & \ra{19}{01}{48}{1} & \dec{-36}{57}{21}{9} & embedded  & $\leq$4.2E-16 & ISAAC \\
R~CrA~IRS~5B             & \ra{19}{01}{48}{1} & \dec{-36}{57}{21}{9} & embedded  & $\leq$1.7E-16 & ISAAC \\ 
R~CrA~IRS~5A+B           & \ra{19}{01}{48}{0} & \dec{-36}{57}{21}{6} & embedded  & -             & Spitzer \\ 
HH~100~IRS               & \ra{19}{01}{50}{7} & \dec{-36}{58}{09}{6} & embedded  & $\leq$8.8E-15 & ISAAC \\ 
R~CrA~IRS~7A             & \ra{19}{01}{55}{3} & \dec{-36}{57}{22}{0} & embedded  & $\leq$1.5E-16 & ISAAC, Spitzer \\ 
R~CrA~IRS~7B             & \ra{19}{01}{56}{4} & \dec{-36}{57}{28}{1} & embedded/borderline  & $\leq$7.1E-17 & ISAAC, Spitzer \\ 
R~CrA~IRAS32             & \ra{19}{02}{58}{7} & \dec{-37}{07}{34}{7} & embedded  & -             & Spitzer \\
\hline
Additional sources \\
\hline
IRAS~13546-3941         & \ra{13}{57}{38}{9} & \dec{-39}{56}{00}{2} & disk  & -             & Spitzer \\ 
IRAS~15398-3359         & \ra{15}{43}{02}{3} & \dec{-34}{09}{06}{8} & embedded  & -             & Spitzer \\ 
IRAS~23238+7401         & \ra{23}{25}{46}{7} & \dec{+74}{17}{37}{3} & embedded  & -             & Spitzer \\ 
\end{longtable}
} % end \longtab
} % end scriptsize
%

%###############
\bibliographystyle{aa}
\bibliography{11001}
%###############

 \end{document}